\documentclass[aps,amsmath,amssymb,twocolumn,prb,showpacs]{revtex4}

\usepackage{graphicx}
\usepackage{color}

\begin{document}


\title{Near-field intensity correlations in parametric photoluminescence from a planar microcavity}

\author{Davide Sarchi}
\email[]{sarchi@science.unitn.it}
\author{Iacopo Carusotto}
\affiliation{CNR-INFM BEC Center and Dipartimento di Fisica,
Universit\`a di Trento, via Sommarive 14, I-38050 Povo, Trento,
Italy}

\date{\today}
\begin{abstract}
We study the spatio-temporal pattern of the near-field intensity
correlations generated by parametric scattering processes in a
planar optical cavity. A generalized Bogolubov-de Gennes model is
used to compute the second order field correlation function.
Analytic approximations are developed to understand the numerical
results in the different regimes. The correlation pattern is found
to be robust against a realistic disorder for state-of-the-art
semiconductor systems.
\end{abstract}
\pacs{71.36.+c, 42.50.Ar, 42.65.Yj, 71.35.Lk}
\maketitle                   
\section{Introduction}

Quantum correlations in many-body and optical systems are playing a
crucial role in a variety of fields, from the microscopic study of
novel states of matter in quantum
fluids~\cite{stringaribook,bloch08}, to the control and suppression
of noise in optical systems
\cite{lugiato_review,gatti09,zambrini00,lopez08,delaubert08}, to the
exploration of analog models of gravitational and cosmological
systems \cite{balbinot08,carusotto08,carusotto09}.

In this perspective, non-linear optical systems exhibit a rich
variety of phenomena~\cite{NLO,milburn94}, involving  the interplay
of parametric scattering with losses, non-linear energy shifts, and
the peculiar dispersion of light in confined geometries. As a most
significant example of this physics, an example of phase transition
of the Bose-Einstein class has recently started being investigated
in the optical parametric oscillation in planar geometries
\cite{savvidis00,ciuti03}.

Theoretical work has suggested that further insight in this physics
can be obtained from the spatial and temporal pattern of
correlations of the emitted light~\cite{carusotto05,drummond05}: as
the critical point for parametric oscillation is approached, the
correlation length and time of the parametric emission show a
divergence that is closely related to the one of thermodynamical
phase transitions.

To this purpose, semiconductor microcavities in the strong coupling
regime appear as most favorable
systems~\cite{savvidis00,ciuti03,deveaud_specissue} as they are
intrinsically grown with a planar geometry, and nonlinear
interactions between the dressed photons -- the so-called polaritons
-- are remarkably strong. First experiments demonstrating quantum
correlations have recently appeared
\cite{langbeinprl,giacobino06,giacobino07}.

In the present work, we investigate the physics of the intensity
correlations that are generated by parametric scattering processes.
Depending on the pump frequency and intensity, different regimes can
be identified. In addition to the usual short-distance correlations
that are generally present in any interacting system, spontaneous
parametric emission processes are responsible for additional
long-distance ones, whose correlation length diverges as the optical
parametric oscillation threshold is approached. In addition to their
intrinsic interest, an experimental study of intensity correlation
in this simplest system will open the way to the investigation of
more complex geometries that in many aspects mimic the behavior of
quantum fields in curved space-times \cite{marino08}.

Our approach is based on a generalized, non-equilibrium Bogolubov-de
Gennes approximation in which weak fluctuations around the classical
coherent field are described in terms of a quadratic Hamiltonian
\cite{castin_school,ciuti03,verger07,sarchi09}. Solving in frequency
space the corresponding quantum Langevin equations allows to obtain
predictions for the spatial and temporal dependence of the in-cavity
intensity correlations. These then directly transfer to the
near-field correlations of the emitted light. The study of the
simplest planar geometry is extended to the more realistic case of
weakly disordered systems: in this regime, the effect of disorder is
shown not to qualitatively modify the peculiar correlation pattern.

In Section \ref{sec:theory}, we review the theoretical formalism
based on the Bogolubov approximation and we discuss how correlations
transfer from the in-cavity field to the emitted radiation. In
Section \ref{sec:numres}.A we present the numerical results for a
one-dimensional, spatially uniform case. Approximate analytical
calculations are presented in Section \ref{sec:numres}.B and used to
physically interpret the numerical results. Numerical results for a
two-dimensional disordered system are discussed in Section
\ref{sec:dis}. Conclusions are drawn in Section \ref{sec:concl}.

\section{Formalism}
\label{sec:theory}

In this section we briefly review the formalism which will be used
to calculate the intensity correlations of the emitted light.  As we
are restricting to the case of small fluctuations around a strong
coherent field, a Bogolubov approach is adapted to describe the
dynamics of quantum fluctuations. In the simplest case of a
spatially homogeneous geometry the Bogolubov equations can be worked
out to analytical expressions for the physical observables. In the
general case, numerical results can be obtained by inverting the
Bogolubov matrix for the specific geometry under investigation.

\subsection{The system Hamiltonian}

We describe the exciton and the photon quantum fields
$\hat{\Psi}_{x}({\bf r})$ and $\hat{\Psi}_{c}({\bf r})$ as scalar
Bose fields. The Hamiltonian is the sum of terms describing the free
propagation of excitons and photons $\hat{H}_0$, their dipole
coupling $\hat{H}_R$, the two-body exciton-exciton interaction
$\hat{H}_X$, saturation of the exciton-photon coupling $\hat{H}_s$
and the external pumping $\hat{H}_p$~\cite{ciuti03,carusotto04}:
\begin{equation}
\hat{H}=\hat{H}_{0}+\hat{H}_{R}+\hat{H}_{x}+\hat{H}_{s}+\hat{H}_p.
\label{eq:Hcomp}
\end{equation}
The free propagation term has the form
\begin{eqnarray}
\hat{H}_{0}&=&\int d^2{\bf r}\hat{\Psi}^{\dagger}_{x}({\bf r})[E_x-(\hbar^2/2m_x)\nabla^2+U_{x}({\bf r})]\hat{\Psi}_{x}({\bf r})\nonumber\\
&+&\int d^2{\bf r}\hat{\Psi}^{\dagger}_{c}({\bf
r})[\epsilon_c(-i\vec{\nabla})+U_{c}({\bf r})]\hat{\Psi}_{c}({\bf
r}) \label{eq:H0}
\end{eqnarray}
where $E_x$ is the exciton energy, $m_{x}$ its effective mass, and $\epsilon_c$ is the photon dispersion in the planar cavity; $U_{x,c}$ are the external potentials acting on respectively the photon and the exciton.
The dipole exciton-photon coupling has the form
\begin{equation}
\hat{H}_{R}=\hbar\Omega_{R}\int d^2{\bf
r}[\hat{\Psi}^{\dagger}_{x}({\bf r})\hat{\Psi}_{c}({\bf
r})+\textrm{h.c.}] \label{eq:HR}
\end{equation}
and is quantified by the Rabi frequency $\Omega_R$.
The effective two-body exciton-exciton interaction term
\begin{equation}
\hat{H}_{x}=\frac{1}{2}g_x\int d^2{\bf
r}\hat{\Psi}^{\dagger}_{x}({\bf r})\hat{\Psi}^{\dagger}_{x}({\bf
r})\hat{\Psi}_{x}({\bf r})\hat{\Psi}_{x}({\bf r}) \label{eq:Hx}
\end{equation}
models both Coulomb interaction and the effect of Pauli
exclusion on electrons and holes \cite{rochat00,bentabou01}.
\begin{equation}
\hat{H}_{s}=g_{s}\int d^2{\bf r}[\hat{\Psi}^{\dagger}_{c}({\bf
r})\hat{\Psi}^{\dagger}_{x}({\bf r})\hat{\Psi}_{x}({\bf
r})\hat{\Psi}_{x}({\bf r})+\textrm{h.c.}] \label{eq:Hs}
\end{equation}
is the term modeling the saturation of the exciton oscillator
strength \cite{rochat00}.
The external pump term has the form
\begin{equation}
\hat{H}_{p}= \hbar\int d^2{\bf r}[\hat{\Psi}^{\dagger}_{c}({\bf
r})\,F_p({\bf r},t)+\textrm{h.c.}]\,.
\end{equation}
In the following of the paper, we assume that the system is driven
by a continuous-wave monochromatic pump with a plane wave spatial
profile
\begin{equation}
F({\bf r},t)= e^{-i\omega_p t}e^{i{\bf k}_p \cdot{\bf r}}\,F_0.
\end{equation}
Excitons and photons decay in time with a rate $\gamma_{x}$ and $\gamma_c$, respectively.

\subsection{Bogolubov-de Gennes formalism}

To make the problem analytically tractable, we perform the Bogolubov
approximation~\cite{castin_school}: the two quantum fields are split
into a strong coherent --classical-- component and weak quantum
fluctuations
\begin{equation}
\hat{\Psi}_{x(c)}({\bf r},t)=e^{-i\omega_p t}\left[\Phi_{x(c)}({\bf
r})+\delta\hat{\psi}_{x(c)}({\bf r},t)\right]. \label{eq:bogans}
\end{equation}
This decomposition is then inserted into the Hamiltonian
(\ref{eq:Hcomp}) and all terms of third and higher order in the
fluctuations are neglected. This leads to a quadratic Hamiltonian
for the fluctuation field that can be attacked with available
theoretical tools.

The classical component ${\Phi}_{x(c)}({\bf r})$ is obtained from
the Heisenberg equations of motion of the quantum field
$\hat{\Psi}_{x(c)}({\bf r})$ by factorizing out the multi-operator
averages. This leads to the following pair of generalized
Gross-Pitaevskii equations:
\begin{eqnarray}
\hbar\omega_p\Phi_x({\bf r})&=&\left(-\frac{\hbar^2\nabla^2}{2 m_x}+U_{x}({\bf r})-i\hbar\gamma_x/2+g_x\mid \Phi_x({\bf r})\mid^2\right.\nonumber\\
&+&\left.2g_s\mbox{Re}\left\{\Phi_x^*({\bf r})\Phi_c({\bf r})\right\}\right)\Phi_x({\bf r})\nonumber\\
&+&\left(\hbar\Omega_R+g_s\mid \Phi_x({\bf r})\mid^2\right)\Phi_c({\bf r}), \label{eq:GP_X}\\
\hbar\omega_p\Phi_c({\bf r})&=&\left[\epsilon_c(-i\vec{\nabla})+U_{c}({\bf r})-i\hbar\gamma_c/2\right]\Phi_c({\bf r})\label{eq:GP_C}\\
&+&\left(\hbar\Omega_R+g_s\mid \Phi_x({\bf
r})\mid^2\right)\Phi_x({\bf r})+\hbar F_0 e^{i{\bf k}_p\cdot {\bf
r}}\nonumber \,,
\end{eqnarray}
Note that the classical field $\Phi_{c,x}({\bf r})$ that is obtained
from this equation does not include the correction due to the
backaction of fluctuations onto the coherent
component~\cite{castin97}. To take this effect into account, one
should include the contribution $\langle
\delta\hat{\psi}^\dagger_{x}({\bf r})\, \delta\hat{\psi}_{x}({\bf
r})\rangle$ of the fluctuating fields to the density and then
iteratively solve (\ref{eq:GP_X}-\ref{eq:GP_C}) up to convergence.
As in the present paper we are interested in the correlation
properties of the field fluctuations, this effect can be safely
neglected in what follows.

Within the input-output formalism \cite{milburn94}, the quantum dynamics of fluctuations can be
written in terms of quantum Langevin equations of the form~\cite{ciuti06,verger07,sarchi09}
\begin{equation}
i\hbar\partial_t{\bf \delta \Psi}({\bf r},t)=\hat{M}\,{\bf \delta
\Psi}({\bf r},t)+\hbar {\bf f}({\bf r},t)\,, \label{eq:QLang}
\end{equation}
where
\begin{equation}
{\bf \delta \Psi}=(\delta\psi_x, \delta\psi_x^{\dagger},
\delta\psi_c, \delta\psi_c^{\dagger})^T
\end{equation}
is the four-component quantum fluctuation field. The matrix $\hat{M}$ is obtained by linearizing the Heisenberg equation of motion for the quantum field around the classical component. In our case, it has the form:
\begin{equation}
\hat{M}=\left(\begin{array}{cccc}
\hat{T}_x-i\frac{\hbar\gamma_x}{2} & \Sigma^{xx}_{12} & \tilde{\Omega}_R & \Sigma^{xc}_{12}\\
-(\Sigma^{xx}_{12})^* & -\hat{T}_x-i\frac{\hbar\gamma_x}{2} & -(\Sigma^{xc}_{12})^* & -\tilde{\Omega}_R \\
\tilde{\Omega}_R & \Sigma^{xc}_{12} & \hat{T}_c-i\frac{\hbar\gamma_c}{2} & 0 \\
-(\Sigma^{xc}_{12})^* & -\tilde{\Omega}_R & 0 &
-\hat{T}_c-i\frac{\hbar\gamma_c}{2}\end{array}\right)\,,
\label{eq:MBdG}
\end{equation}
with
\begin{eqnarray}
\hat{T}_x({\bf r})&=&-\frac{\hbar^2\nabla^2}{2 m_x}+U_{x}({\bf r})-\hbar\omega_p+2g_x\mid \Phi_x({\bf r})\mid^2\nonumber\\
&+&4g_s\mbox{Re}\{\Phi_x^*({\bf r})\Phi_c({\bf r})\}\,, \\
\hat{T}_c({\bf r})&=&\epsilon_c(-i\vec{\nabla})+U_{c}({\bf
r})\,,\\
\Sigma_{12}^{xx}({\bf r})&=&g_x\Phi_x^2({\bf r})+2g_s\Phi_x({\bf r})\Phi_c({\bf r})\,, \\
\tilde{\Omega}_R({\bf r})&=&\hbar\Omega_R+2g_s\mid \Phi_x({\bf r})\mid^2\,,\\
\Sigma^{xc}_{12}({\bf r})&=&g_s\Phi^2_x({\bf r})\,.\label{eq:MBdG2}
\end{eqnarray}

The quantum Langevin force
\begin{equation}
{\bf f}=(f_x, f_x^{\dagger}, f_c,
f_c^{\dagger})^T
\end{equation}
describes the zero-point fluctuations in the input field.
As the field dynamics takes place in a small frequency window around $\omega_p$ and the input field is assumed to be in the vacuum state, the spectrum of quantum Langevin force can be approximated by a white noise in both space and time,
\begin{equation}
\langle f_{\xi}({\bf r},t) f^{\dagger}_{\chi}({\bf
r}^{\prime},t')\rangle= \gamma_{\chi} \delta_{\xi,\chi}\,\delta({\bf
r}-{\bf r}^{\prime})\,\delta(t-t').
\end{equation}
This approximation is accurate in the present case where no other
radiation is incident onto the cavity in addition to the driving at
$\omega_p$ and the light-matter coupling $\hbar \Omega_R$ is small
as compared to the natural oscillation frequencies $E_x$ and
$\epsilon_c$~\cite{deliberato07,deliberato09}.

\subsection{Evaluation of the correlation functions}

In the present case of a monochromatic pump, temporal homogeneity guarantees that the correlation functions only depend on the time difference $t-t'$ and different frequency components are decoupled:
\begin{equation}
\hbar\omega\,{\bf \delta\Psi}({\bf r},\omega)=\hat{M}{\bf
\delta\Psi}({\bf r},\omega)+\hbar {\bf f}({\bf r},\omega)\,.
\label{eq:QLang_omega}
\end{equation}
The two-operator averages of the fluctuation fields can be written as
\begin{equation}
\tilde{g}^{(1)}_{x(c)}({\bf r},t;{\bf r'},t') =\int d\omega
e^{-i\omega(t-t')}\tilde{g}^{(1)}_{x(c)}({\bf r},{\bf r'},\omega)
\,, \label{eq:timecorr1}
\end{equation}
and
\begin{equation}
\tilde{m}^{(1)}_{x(c)}({\bf r},t;{\bf r'},t') =\int d\omega
e^{-i\omega(t-t')}\tilde{m}^{(1)}_{x(c)}({\bf r},{\bf r'},\omega)
\,. \label{eq:timecorr2}
\end{equation}
where we have defined
\begin{equation}
\tilde{g}^{(1)}_{x(c)}({\bf r},{\bf r'},\omega) =\int
d\omega^{\prime}\langle\delta\psi_{x(c)}^{\dagger}({\bf
r},\omega)\delta\psi_{x(c)}({\bf r'},\omega^{\prime}) \rangle\,,
\label{eq:freqcorr1}
\end{equation}
\begin{equation}
\tilde{m}^{(1)}_{x(c)}({\bf r},{\bf r'},\omega) =\int
d\omega^{\prime}\langle\delta\psi_{x(c)}({\bf
r},\omega)\delta\psi_{x(c)}({\bf r'},\omega^{\prime}) \rangle\,.
\label{eq:freqcorr2}
\end{equation}
By introducing the frequency-domain correlation functions
\begin{equation}
\langle f_{\xi}({\bf r},\omega) f^{\dagger}_{\chi}({\bf
r}^{\prime},\omega^{\prime})\rangle_{\bf f}=2\pi\gamma_{\chi}
\delta_{\xi,\chi}\delta({\bf r}-{\bf
r}^{\prime})\delta(\omega-\omega^{\prime}) \label{eq:corrfomega}
\end{equation}
and making use of (\ref{eq:QLang_omega}), simple expressions for the two-operator averages are obtained
\begin{equation}
g_{x}({\bf r},{\bf r'},\omega)=\sum_{l=1}^{4}\int d{\bf
s}\langle{\bf s},l|\hat{M}_{\omega}^{-1}| {\bf r'},1\rangle^*
\langle{\bf s},l|\hat{M}_{\omega}^{-1}| {\bf r},1\rangle \Gamma_l\,,
\label{eq:PLres_x}
\end{equation}
\begin{equation}
g_{c}({\bf r},{\bf r'},\omega)=\sum_{l=1}^{4}\int d{\bf
s}\langle{\bf s},l|\hat{M}_{\omega}^{-1}| {\bf r'},3\rangle^*
\langle{\bf s},l|\hat{M}_{\omega}^{-1}| {\bf r},3\rangle \Gamma_l\,,
\label{eq:PLres_c}
\end{equation}
\begin{equation}
m_{x}({\bf r},{\bf r'},\omega)=\sum_{l=1}^{4}\int d{\bf
s}\langle{\bf s},l|\hat{M}_{\omega}^{-1}| {\bf r'},2\rangle^*
\langle{\bf s},l|\hat{M}_{\omega}^{-1}| {\bf r},1\rangle \Gamma_l\,,
\label{eq:CORRres_x}
\end{equation}
\begin{equation}
m_{c}({\bf r},{\bf r'},\omega)=\sum_{l=1}^{4}\int d{\bf
s}\langle{\bf s},l|\hat{M}_{\omega}^{-1}| {\bf r'},4\rangle^*
\langle{\bf s},l|\hat{M}_{\omega}^{-1}| {\bf r},3\rangle \Gamma_l\,.
\label{eq:CORRres_c}
\end{equation}
where ${\bf \Gamma}=2\pi \hbar^2(0,\gamma_{x},0,\gamma_{c})^T$ and the matrix
\begin{equation}
\hat{M}_{\omega}=\hat{M}-\hbar\omega\,{\bf 1}
\end{equation}
is evaluated in the basis $|({\bf r},j)\rangle$. Here, $|{\bf
r}\rangle$ spans the real space and the label $j=1,...,4$ refers to
the block form of the Bogolubov matrix (\ref{eq:MBdG}).

\subsection{Intensity correlations}

The present paper is focussed on the spatio-temporal correlations of
the intensity fluctuations of the in-cavity photon field. As usual
in generic planar microcavities, the in-cavity intensity
correlations directly reflect into the near-field intensity
correlation of the emitted light outside the cavity. An explicit
calculation of the relation between the in-cavity and the emitted
fields in the specific case of semiconductor DBR microcavities is
given in the Appendix.

\begin{center}
\begin{figure*}[htbp]
\includegraphics[clip,width=0.61\textwidth]{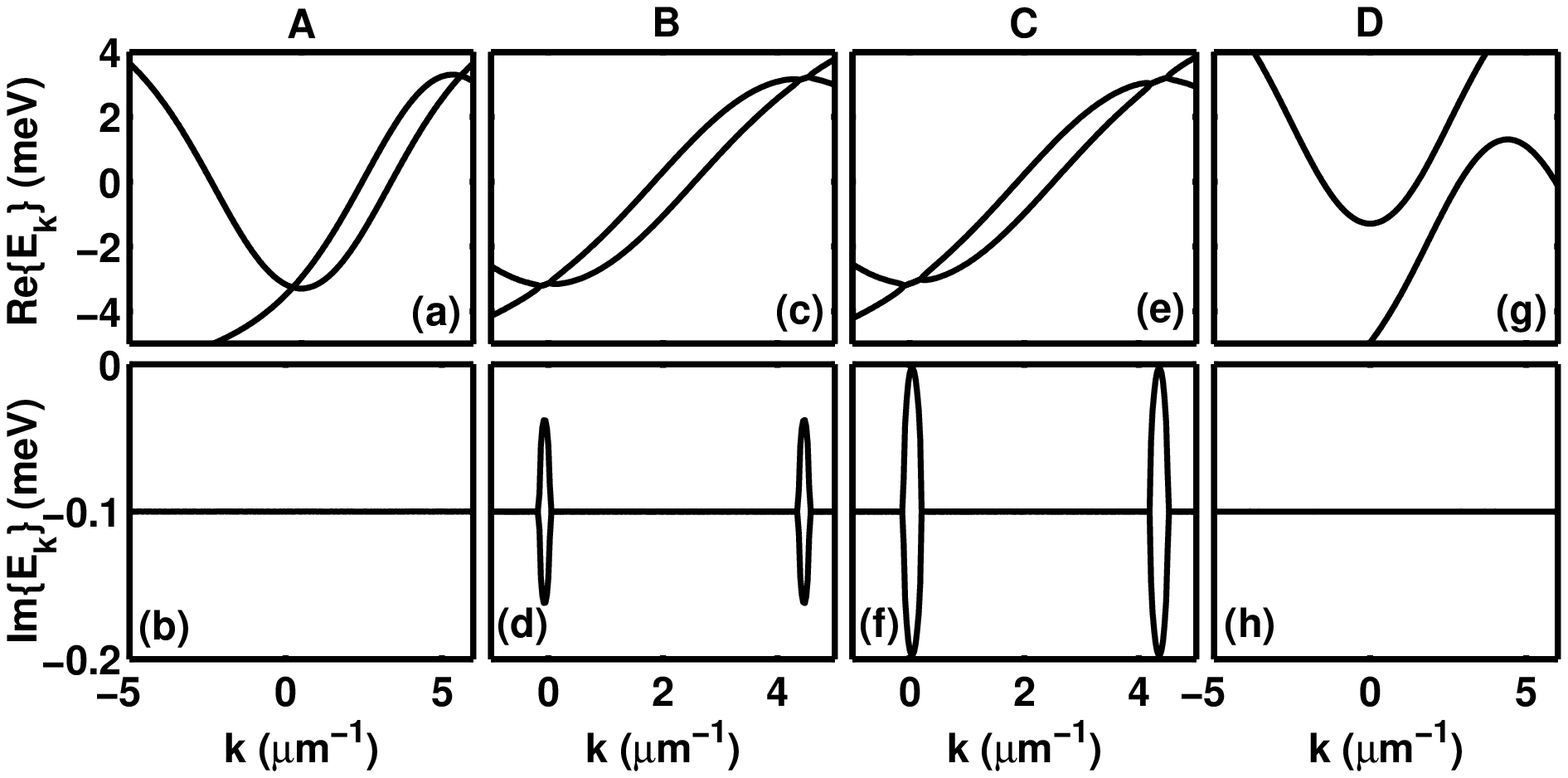}
\includegraphics[width=.365 \textwidth]{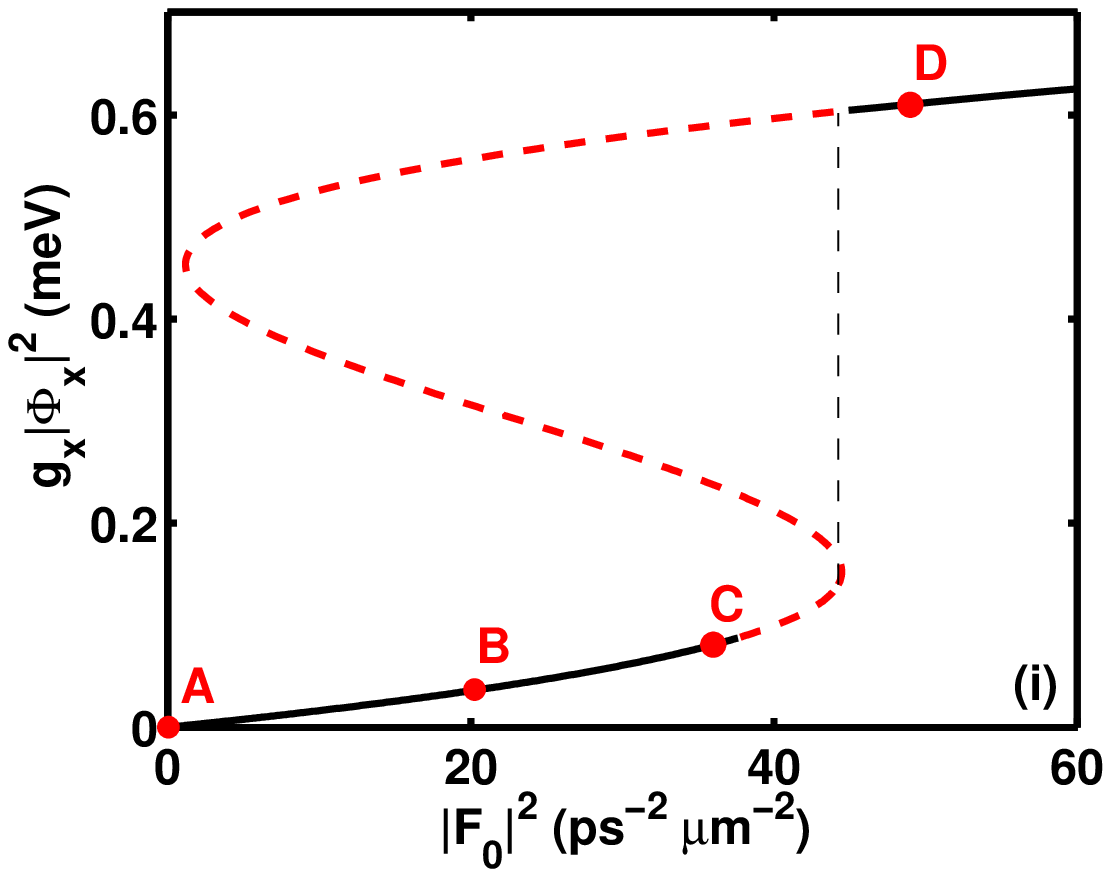}
\caption{Real (a-g) and imaginary parts (b-h) of the Bogolubov
dispersion as a function of the wavevector $k$, for the four regimes
A-D marked in panel (i). The exciton interaction energy is
$g|\Phi_x|^2=10^{-4}$ meV (a, b), $g|\Phi_x|^2=0.04$ meV (c, d),
$g|\Phi_x|^2=0.08$ meV (e, f) and $g|\Phi_x|^2=0.61$ meV (g, h). (i)
Bistable loop in the in-cavity exciton field intensity as a function
of the incident pump intensity. The red dashed line represents the
unstable solutions. The points labeled with the letters {\bf A}-{\bf
D} mark the four regimes considered in panels (a-h).}
\label{fig:disp}
\end{figure*}
\end{center}

The four operator correlation function describing the correlation of the intensity fluctuations of the in-cavity photon field is defined as usual as:
\begin{equation}
G^{(2)}({\bf r},t; {\bf r'},t')=\langle \hat{\Psi}_c^{\dagger} ({\bf
r},t)\hat{\Psi}_c^{\dagger}({\bf r'},t') \hat{\Psi}_c({\bf r'},t')
\hat{\Psi}_c({\bf r},t) \rangle \,.\label{eq:G2}
\end{equation}
At the level of the Bogolubov approximation, three- and
four-operator averages of the fluctuation field can be neglected and
the correlation function (\ref{eq:G2}) can be written in terms of
two-operator averages as
\begin{eqnarray}
G^{(2)}({\bf r},{\bf r'},t)&=&|\Phi_c({\bf r})|^2 |\Phi_c({\bf
r'})|^2+|\Phi_c({\bf r'})|^2\langle \hat{\psi}_c^{\dagger} ({\bf
r})\hat{\psi}_c({\bf r}) \rangle\nonumber \\&+&|\Phi_c({\bf
r})|^2\langle \hat{\psi}_c^{\dagger} ({\bf r'})\hat{\psi}_c({\bf
r'}) \rangle \nonumber \\&+&2\mbox{Re}\{\Phi_c({\bf
r'})^*\Phi_c({\bf r}) \langle \hat{\psi}_c^{\dagger} ({\bf
r},t)\hat{\psi}_c({\bf r'},0) \rangle \}\nonumber
\\&+&2\mbox{Re}\{\Phi_c({\bf r'})^*\Phi_c({\bf r})^* \langle
\hat{\psi}_c ({\bf r},t)\hat{\psi}_c({\bf r'},0) \rangle \}
\,.\label{eq:G2expr}
\end{eqnarray}
A particularly useful quantity is the reduced correlation function:
\begin{equation}
\bar{G}^{(2)}({\bf r},t;{\bf r'},t')=\frac{G^{(2)}({\bf r},t;{\bf
r',t'})}{n_c({\bf r})\,n_c({\bf r'})}-1\,, \label{eq:G2red}
\end{equation}
scaled by the intensity of light
\begin{equation}
n_c({\bf r})=|\Phi_c({\bf r})|^2+\langle \delta\hat{\psi}^\dagger_c({\bf r})\,
\delta\hat{\psi}_c({\bf r})\rangle \,,
\label{eq:G2red_2}
\end{equation}
which can be written in the following simple form:
\begin{eqnarray}
\bar{G}^{(2)}({\bf r},{\bf r'},t)&\simeq&2\mbox{Re}\left\{\Phi_c({\bf
r'})^*\Phi_c({\bf r}) \langle \hat{\psi}_c^{\dagger} ({\bf
r},t)\hat{\psi}_c({\bf r'},0) \rangle \right.\nonumber
\\&+&\left.\Phi_c({\bf r'})^*\Phi_c({\bf r})^* \langle
\hat{\psi}_c ({\bf r},t)\hat{\psi}_c({\bf r'},0) \rangle
\right\}\times\nonumber\\
&\times&[|\Phi_c({\bf r})|^2|\Phi_c({\bf r'})|^2]^{-1}
\,.\label{eq:G2rexpr}
\end{eqnarray}
In the next section, we will study the spatial and temporal features of this quantity
in the most relevant cases.

\section{1D uniform system}
\label{sec:numres}

\subsection{Numerical results}

In this section we investigate the case of a spatially uniform
system under a plane wave monochromatic pump. Spatial homogeneity
guarantees that the correlation functions only depend on the
relative spatial coordinate $r=x-x'$.

We use typical parameters for a GaAs microcavity with $N=10$ quantum
wells. In particular, we choose $\hbar \Omega_R=10$ meV,
$\hbar\gamma_c=\hbar\gamma_x=\hbar\gamma=0.1$ meV, and we take zero
exciton-photon detuning at $k=0$, $E_x=\epsilon_c(k=0)$. For the
sake of simplicity, in the present section we focus our attention on
a quasi-one-dimensional geometry where polaritons are transversally
confined to a length $l_{1D}\sim 1~\mu$m. This corresponds to
effective 1D nonlinear coupling constants $g_x=1.5\times 10^{-3}$meV
$\mu$m and $g_s=0.5\times 10^{-3}$ meV $\mu$m. Such geometries are
presently under active experimental investigation \cite{dasbach05}.
Extension to the 2D case does not qualitatively affect the physics
and will be discussed in Sec. \ref{sec:dis} in connection to
disorder issues.

We consider a configuration where the pump wave vector
$k_p=2.2~\mu$m$^{-1}$ is close to the magic wave vector, and the
pump energy is $\hbar\omega_p-\epsilon_c(0)=-8.7$ meV. This
configuration allows to study the most significant regimes by simply
varying the pump intensity. In Fig. \ref{fig:disp} (i), we plot the
corresponding exciton field intensity as a function of the pump
intensity $|F_0|^2$: the hysteresis cycle typical of bistable
systems is apparent~\cite{baas04_bistab,carusotto04,ciuti_pssb}.

In the following, we will consider four different density regimes,
marked on the bistability loop by the points {\bf A}-{\bf D}. Two of
them ({\bf A}, {\bf B}) correspond to low-density conditions far
from any instability, the point {\bf C} is very close to the OPO
threshold, and point {\bf D} is in the stable region above the
bistability and parametric oscillation region.

In the present uniform system, the elementary excitations on top of
the steady-state of the pumped system can be classified in terms of
their wavevector $k$. Examples of their Bogolubov
dispersion~\cite{carusotto04} are shown in Fig. \ref{fig:disp} (a-h)
for the four different regimes marked in Fig. \ref{fig:disp} (i) by
the points {\bf A}-{\bf D}. In the figure, we restrict the field of
view to the lower-polariton region that is involved in the physics
under investigation here.

For increasing intensities, we observe: i) the low-density
parametric regime {\bf A} (panels a, b), where the Bogolubov modes
reduce to the single-particle dispersion; ii) the regime {\bf B}
corresponding to moderate densities (panels c, d), where the
imaginary parts are modified and a small region of flattened
dispersion appears at the crossing of the Bogolubov modes; iii) the
regime {\bf C} close to OPO threshold (panels e, f), where the
imaginary part of one mode tends to zero and a flat region in the
Bogolubov dispersion is apparent; iv) the non-parametric
configuration {\bf D}, for intensities larger than the bistability
threshold, where the normal and the ghost dispersions are well
separated and the eigenmodes of the system tend again to the
single-particle dispersion, yet blue-shifted by interactions (panels
g, h).

For each of this {\bf A}-{\bf D} regimes, we have computed the
spatio-temporal pattern of intensity correlations with the model
described in the previous section. The results are presented in the
next subsections. An analytical interpretation will be given later
on in Sec. \ref{sec:numres}.B.

\subsubsection{Low-density parametric luminescence}

\begin{figure}[htbp]
\includegraphics[clip,width=.48 \textwidth]{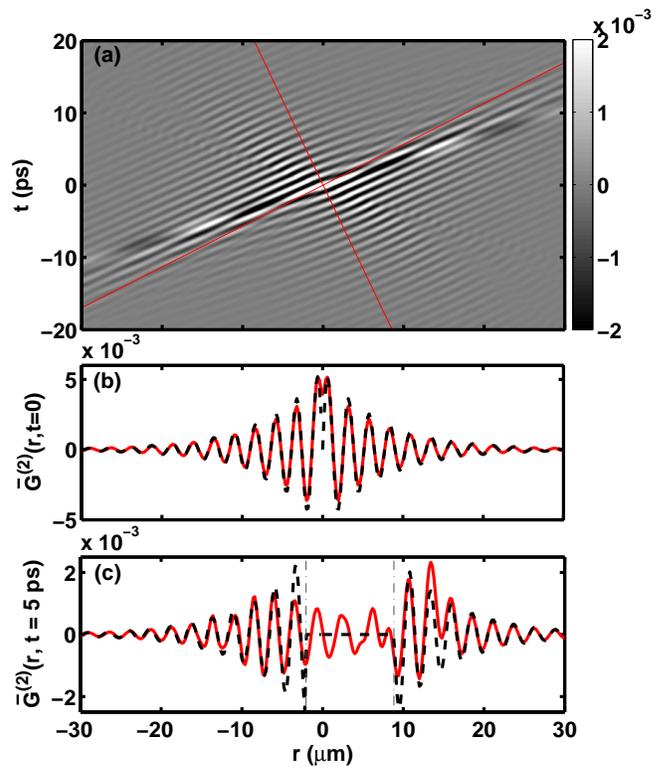}
\caption{(a) Pattern of the reduced intensity correlations
$\bar{G}^{(2)}(r,t)$ as a function of the spatial distance $r$ and
of the temporal delay $t$, for the low-density regime A,
corresponding to Fig. \ref{fig:disp} (a, b). The two thin lines,
$r=v_g^s t$ and $r=v_g^i t$, highlight the group velocities of the
signal and idler modes. (b) Cut of the figure (a) along the zero
delay $t=0$ line. (c) Cut along the $t=5$ ps line. In panels (b) and
(c), the analytical prediction Eq. (\ref{eq:G2kdom_LD2}) is shown as
a dashed line. In panel (c), the two vertical dash-dotted lines
delimit the region where the analytical result (\ref{eq:G2kdom_LD2})
predicts vanishing correlations.} \label{fig:LD}
\end{figure}

In Fig. \ref{fig:LD} (a), we show the spatio-temporal pattern of the
intensity correlation in the low-density regime A corresponding to
Fig. \ref{fig:disp} (a, b): this is characterized by a system of
parallel fringes and a butterfly-shaped envelope. The fringe
amplitude is almost vanishing inside the cone delimited by the group
velocities of the signal and the idler modes (represented by the
thin solid lines in the figure) and largest on the edges of the
cone. Further away in space and time, it decays back to zero as a
consequence of the finite polariton lifetime. At zero delay $t=0$,
the central bunching peak around $r=0$ shows a weak narrow dip
[Fig.\ref{fig:LD}(b)]. At larger delays [Fig.\ref{fig:LD}(c)], the
correlation signal is weaker in between the dot-dashed lines
indicating the edges of the cone.

\subsubsection{Moderate-density parametric luminescence}

\begin{figure}[htbp]
\includegraphics[clip,width=.48 \textwidth]{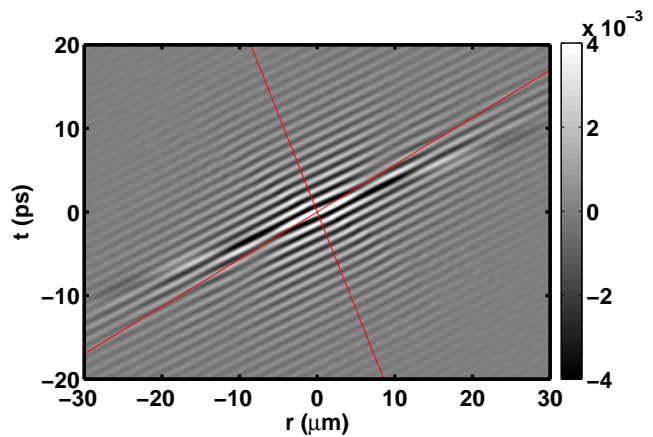}
\caption{(a) Pattern of the intensity correlations
$\bar{G}^{(2)}(r,t)$, for the moderate density regime B, still well
below the OPO threshold and corresponding to Fig. \ref{fig:disp} (c,
d). The two thin lines, $r=v_g^s t$, and $r=v_g^i t$, highlight the
group velocities of the signal and idler modes.} \label{fig:MD}
\end{figure}

In Fig. \ref{fig:MD} (a), we show the spatio-temporal pattern in the
moderate density regime B corresponding to the energy dispersion
shown in Fig. \ref{fig:disp} (c,  d). As compared to the low-density
case, the qualitative shape of the correlation pattern is
qualitatively modified: the system of parallel fringes extends in a
significant way into the interior of the butterfly shape and the
exponential decay in the external region takes place at a slower
rate. This latter effect is a direct consequence of the increased
lifetime of the Bogolubov modes in the parametric region (see Fig.
\ref{fig:disp} (d)).

\subsubsection{OPO critical region}

\begin{figure}[htbp]
\includegraphics[clip,width=.48 \textwidth]{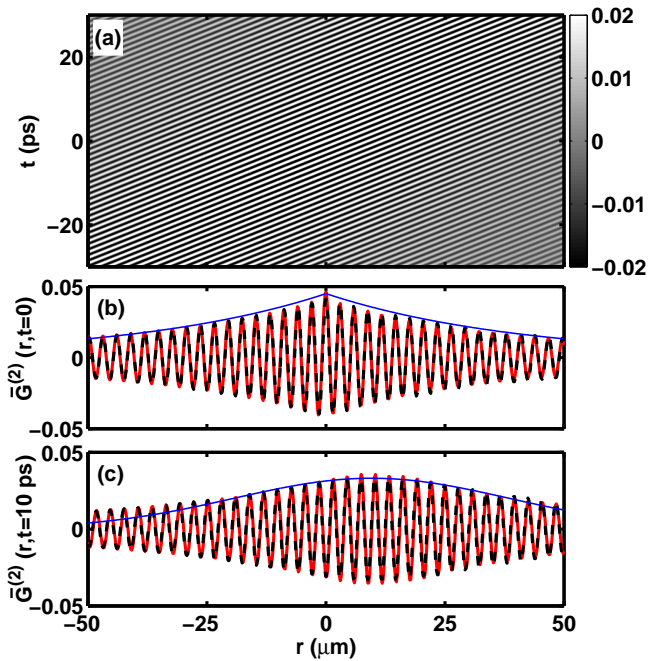}
\caption{(a) Pattern of the intensity correlations
$\bar{G}^{(2)}(r,t)$ for the density regime C just below the OPO
threshold and corresponding to Fig. \ref{fig:disp} (e, f). (b) Cut
of the figure (a) along the zero delay $t=0$ line. (c) Cut along the
$t=10$ ps line. In panels (b) and (c), the dashed line is the result
of the analytical prediction Eqs. (\ref{eq:G2kdom1_HD}) and
(\ref{eq:G2kdom_HD}).} \label{fig:HD}
\end{figure}

In Fig. \ref{fig:HD} (a), we show the spatio-temporal pattern for
the regime C very close to the OPO threshold. In this case, the
system of parallel fringes extends to the whole $(r,t)$ space and
correlations are non-vanishing even at very long time and space
separations. This is a consequence of the diverging correlation
length of fluctuations in the critical region~\cite{carusotto05}.
The zero-delay $t=0$ cut of the correlation pattern is shown in Fig.
\ref{fig:HD} (b) and is characterized by a system of fringes
centered at $r=0$ with a central bunching peak. The weak dip that
was visible in the weak intensity regime is no longer present.

\subsubsection{Non-parametric regime}

\begin{figure}[htbp]
\includegraphics[clip,width=.48 \textwidth]{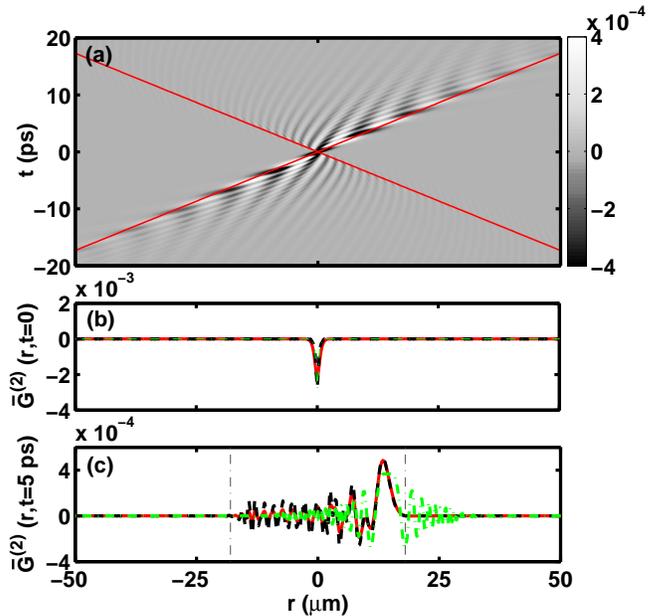}
\caption{(a) Pattern of the intensity correlations
$\bar{G}^{(2)}(r,t)$ for the high-density regime D, corresponding to
Fig. \ref{fig:disp} (g, h). In this regime, the large blue-shift of
the polariton modes prevents the parametric oscillation. The two
thin lines $r=\pm v_g^{max} t$ mark the maximal group velocities.
(b) Cut of the previous figures along the zero delay $t=0$ line. (c)
Cut along the $t=5$ ps line. In panels (b) and (c), the analytical
prediction Eq. (\ref{eq:G2kdom1_NOP1}) (red dashed line) and Eq.
(\ref{eq:G2kdom1_NOP}) (green dash-dotted line) are also shown. In
panel (c), the two vertical thin dash-dotted lines delimit the
region where both the numerical result and the analytical result
(\ref{eq:G2kdom1_NOP1}) predict vanishing correlations.}
\label{fig:NOP}
\end{figure}

For even larger values of the pump intensity beyond the bistability
threshold, the system is pushed into an higher density D regime,
where the blue-shift induced by interactions brings parametric
scattering processes far off-resonance [Fig. \ref{fig:disp} (g, h)].
As a consequence, the system of parallel fringes due to parametric
correlations almost completely disappears, as it can be seen in Fig.
\ref{fig:NOP} (a). The remaining correlation pattern has an opposite
shape, with substantial correlations limited to the region inside
the cone delimited by the maximal group velocity of the polariton
dispersion. At zero delay $t=0$ [Fig. \ref{fig:NOP} (b)] no spatial
fringe is visible and the only feature is an anti-bunching dip at
$r=0$. At finite time delays [Fig. \ref{fig:NOP} (c)], some spatial
fringes appear, but they do not show any dominating periodicity.
This suggests that a continuum of processes at different $k$'s are
simultaneously taking place, each of them characterized by a
different phase velocity.

\subsection{Analytic model and interpretation}
\label{sec:analytic}

To physically understand the numerical results presented in the
previous section, approximate analytic formulas for the behavior of
$\bar{G}^{(2)}$ can be extracted for the most significant limiting
cases. To simplify the analysis, we restrict our attention to the
lower polariton branch. This approximation is accurate as long as
the Rabi splitting is much larger than all other energy scales, e.g.
$\hbar \Omega_R\gg\gamma$, $\hbar\Omega_R\gg g_x|\Phi_x|^2,\,g_s
|\Phi_c\,\Phi_x|$, and is safely fulfilled in all the cases that we
are considering here. In this regime, the dynamics of the system can
be described by a single polariton field. Its dispersion, linewidth
$\gamma$ and nonlinear interaction coefficient $g$ are immediately
obtained from the linear eigenmodes of the Gross-Pitaevskii
equations (\ref{eq:GP_X}) and (\ref{eq:GP_C}) once interactions are
neglected~\cite{ciuti03}. At the same level of approximation, also
the $k$-dependence of the photon Hopfield factor $C_k\simeq C$ can
be neglected. As a result, the same Hopfield coefficient $|C|^4$
appears in both the numerator and in the denominator of the
intensity correlation function (\ref{eq:G2red}) of the photon field,
which then reduces to the corresponding quantity for the
lower-polariton field.

For a spatially homogeneous system under a monochromatic plane wave
pump, equations can be written in the momentum space and the reduced
second-order correlation function reads
\begin{multline}
\bar{G}^{(2)}({\bf r},t)= G^{(2)}_{1}({\bf
r},t)+G^{(2)}_2({\bf r},t)= \\
=\frac{2}{|\Phi|^2}\, \textrm{Re}\left[\int\! \frac{d^d{\bf
k}}{(2\pi)^d}\,e^{-i({\bf k}-{\bf k}_p)\cdot{\bf
r}}\langle\delta\psi^{\dagger}_{\bf k}(t)\delta\psi_{{\bf
k}}(0)\rangle\right]+ \\
+\frac{2}{|\Phi|^4}\,\textrm{Re}\left[\Phi^{*2}\,\int\!
\frac{d^d{\bf k}}{(2\pi)^d}\, e^{i({\bf k}-{\bf k}_p)\cdot{\bf
r}}\langle\delta\psi_{\bf k}(t)\delta\psi_{\bar{\bf
k}}(0)\rangle\right] \label{eq:G2kdom}
\end{multline}
where $\bar{{\bf k}}=2{\bf k}_{p}-{\bf k}$.
In the presence of quantum fluctuations only\cite{ciuti03,verger07}, the normal correlations are
\begin{equation}
\langle\delta\psi^{\dagger}_{\bf k}(t)\delta\psi_{{\bf
k}}(0)\rangle=2\pi\hbar^2\,\gamma\, g^2\,|\Phi|^4 \int d\omega
\frac{e^{i\omega t}}{\Delta_{\bf k}(\omega)}\,,\label{eq:corrNkdom}
\end{equation}
and the anomalous correlations are
\footnote{Note that the anomalous correlations respect the symmetry condition
$\langle\delta\psi_{\bf k}(0)\delta\psi_{\bar{{\bf
k}}}(0)\rangle=\langle\delta\psi_{\bar{{\bf k}}}(0)\delta\psi_{\bf
k}(0)\rangle$. Although it is not evident in Eq.
(\ref{eq:corrkdom}), this property derives from the relations
$E_k^{(1)}-E_k^{(2)}=E_{\bar{k}}^{(1)}-E_{\bar{k}}^{(2)}$,
$E_k^{(1)}-(E_k^{(2)})^*=E_{\bar{k}}^{(1)}-(E_{\bar{k}}^{(2)})^*$,
$E_k^{(1,2)}+\epsilon_{\bar{k}}=E_{\bar{k}}^{(1,2)}+\epsilon_k$,
which highlight the symmetry between the signal and the idler
modes.}
\begin{equation}
\langle\delta\psi_{\bf k}(t)\delta\psi_{\bar{{\bf
k}}}(0)\rangle=-2\pi \hbar^2\,\gamma\, g\, \Phi^2 \int d\omega\,
e^{-i\omega t} \frac{\tilde{\epsilon}_{\bar{\bf k}}+\hbar
\omega+i\frac{\hbar\gamma}{2}}{\Delta_{\bf
k}(\omega)}\,.\label{eq:corrkdom}
\end{equation}
Here we have introduced the notation $\tilde{\epsilon}_{\bf
k}=\epsilon_{\bf k}+2g|\Phi|^2-\hbar\omega_p$ for the shifted
single-particle dispersion of polaritons, ${\bf k}_p$ and $\omega_p$
respectively indicate the wave vector and the frequency of the pump.
The denominator $\Delta_{\bf k}$ is defined in terms of the
eigenvalues $E_{\bf k}^{(1,2)}$ of the Bogolubov matrix
\begin{equation}
M_{\bf k}=\left(\begin{array}{c c}
\tilde{\epsilon}_{\bf k}-i\hbar \gamma/2 & g\Phi^2\\
-g{\Phi^*}^2 & -\tilde{\epsilon}_{\bar{\bf
k}}-i\hbar\gamma/2\end{array}\right)\,, \label{eq:Mmatrix}
\end{equation}
as
\begin{equation}
\Delta_{\bf k}(\omega)=(\hbar\omega-E_{\bf
k}^{(1)})(\hbar\omega-{E_{\bf k}^{(1)}}^*)(\hbar\omega-E_{\bf
k}^{(2)})(\hbar\omega-{E_{\bf k}^{(2)}}^*)\,. \label{eq:denoene}
\end{equation}
To better compare to the numerical results, we now restrict our attention to one-dimensional case and we extract analytic formulas for the most significant regimes considered in the previous subsection.

\subsubsection{Low-density parametric luminescence}

In the low-density limit $|\Phi|\rightarrow 0$ the Bogolubov
eigenmodes tend to the single-particle energies with a negligible
blue-shift [Fig. \ref{fig:disp} (a, b)]. Comparing the expressions
(\ref{eq:corrNkdom}) and (\ref{eq:corrkdom}) for respectively the
normal and the anomalous correlations, it is immediate to see that
the former is a factor $g\,|\Phi|^2/\hbar\gamma$ smaller and
therefore negligible in this limit. By performing the frequency
integral in (\ref{eq:G2kdom}) with the method of residuals, the
spatial correlation function can be written as
\begin{multline}
\bar{G}^{(2)}(r,t)\simeq - 4\pi g\,
\textrm{Sgn}(t)\, e^{-\gamma|t|/2}\times\\
\times \textrm{Re}\left\{\int d k\,
e^{i(k-k_p)r}\,\frac{e^{-i\epsilon_kt/\hbar}}{\epsilon_k+{\epsilon}_{\bar{k}}-2\hbar\omega_p-i\hbar\gamma}\right\}\,
\label{eq:G2kdom_LD1}
\end{multline}

In the parametric configuration (see Fig. \ref{fig:disp} (a, b)),
the dominating contribution to the integral over $k$ is given by the
wavevectors around the signal $k_s$ and the idler
$k_i=\bar{k}_s=2k_p-k_s$ modes, where the normal and the ghost
dispersions intersect and parametric processes are resonant. We can
then approximate the integrand by replacing the energies by their
first-order expansions around $k=k_{s,i}$
\begin{equation}
\epsilon_k\simeq\epsilon_{k_{s,i}}+\hbar v_g^{(s,i)}(k-k_{s,i})\,,
\end{equation}
with group velocities
\begin{equation}
v_g^{(s,i)}=\frac{1}{\hbar}\,\partial_k \epsilon(k)|_{k_{s,i}}\,.
\end{equation}
In this way, we obtain
\begin{multline}
\bar{G}^{(2)}_{2}(r,t)\simeq 8 \pi^2 \frac{g}{\hbar \Delta v_g}
\mbox{Sgn}(t)e^{-\gamma |t|/2}S\sin(K r-\Omega t)\times \\
\times [\theta(t S R_s)e^{-\kappa R_s}-\theta(-t SR_i(r,t))e^{\kappa
R_i(r,t)}] \,, \label{eq:G2kdom_LD2}
\end{multline}
where $S=\mbox{Sgn}(\Delta v_g)$ and $\Delta v_g=v_g^{(s)}-v_g^{(i)}$.

From this expression, it is immediate to see that the system of
parallel fringes has a frequency
$\hbar\Omega={\epsilon}_{k_{s}}-\hbar\omega_p$ and a wavevector
$K=k_s-k_p$ determined by the interference of the signal/idler and
the pump mode. The analytic form of the zero delay $t=0$ fringe
pattern resulting from the combination of $\sin$ and $\theta$
functions is responsible for the dip at $r=0$ that is visible in
Fig.\ref{fig:LD}(b). The temporal decay of the correlation occurs on
the same time scale as the bare polariton decay rate $\gamma$. The
spatial decay away from the butterfly edges (we have set
$R_{s,i}(r,t)=r-v_g^{(s,i)}t$) occurs on a length scale
$\kappa^{-1}=|\Delta v_g|\,/\gamma$.

\begin{figure}[htbp]
\includegraphics[clip,width=.48 \textwidth]{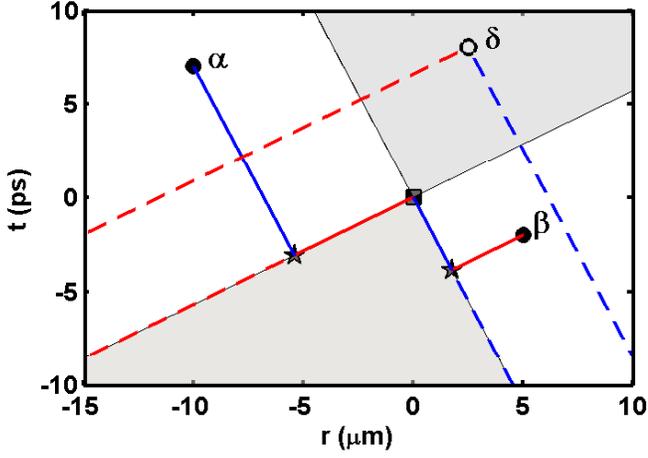}
\caption{Geometric construction to determine the pairs of points in
space-time that are correlated in the low-density parametric
luminescence regime. The shaded area highlights the region of
space-time which is not correlated with the point ($r=0$, $t=0$).
The point ($r=0$, $t=0$) is parametrically correlated with the point
$\alpha$, as a result of the parametric emission event occurring at
the point marked with a star. The same holds for the point $\beta$.
Solid lines mark the motion of signal (blue) and idler (red)
particles. No parametric event can instead produce correlations
between the point ($r=0$, $t=0$) and the point $\delta$ located
inside the shaded region. } \label{fig:geom}
\end{figure}

Furthermore, the analytical formula (\ref{eq:G2kdom_LD2}) shows that
no correlation is present inside the cone marked by the thin lines
in the Fig. \ref{fig:LD} and defined (for $t>0$) by the condition
$R_s<0$ and $R_i>0$ (for $t<0$ the signs are exchanged). The physics
behind this fact is illustrated by the simple geometric construction
shown in Fig. \ref{fig:geom}: pairs of entangled signal/idler
polaritons are generated at all times and positions by the
parametric conversion of quantum fluctuations into real excitations.
Signal and idler polaritons then propagate with group velocities
respectively $v_g^{s,i}$ and transport the correlation to distant
pairs of points. It is easy to see that the shaded region inside the
cone can never be reached by such a process. An analogous reasoning
was used in Ref.~\cite{carusotto08} to explain the correlation
signal observed in numerical calculations of Hawking radiation from
acoustic black holes. As polaritons decay at a rate $\gamma$, the
same construction shows that the correlation signal has to decay in
space with the characteristic length $\kappa^{-1}=\Delta
v_g/\gamma$.

The perfect agreement between this analytic
approximation and the numerical result is highlighted in Fig.
\ref{fig:LD} (b, c). The lower polariton parameters $g$, $\gamma$,
$\epsilon_k$ are calculated from the linear eigenmodes of the GP
equations (\ref{eq:GP_X}) and (\ref{eq:GP_C}). As expected for a
Bogolubov theory~\cite{castin_school}, for a given blue shift
$g\,|\Phi|^2$ the rescaled correlation function is proportional to
the nonlinear coupling constant $g$.

Before proceeding, it is interesting to note that for a pump at
normal incidence $k_p=0$, one has $k_s=-k_i$ and a vanishing
relative frequency $\Omega=0$. The spatial fringes are then
independent on time. From an experimental point of view, this
configuration appears to be the most suitable one to observe the
predicted features as it is less subject to the finite time
resolution of photodetectors.

\subsubsection{OPO critical region}

For pump intensities just below the OPO threshold,
$g|\Phi|^2\lesssim g|\Phi_{OPO}|^2=\hbar\gamma/2$, the Bogolubov
spectrum strongly differs from the single-particle one [Fig.
\ref{fig:disp} (e, f)]. In particular, the imaginary part of one of
the two eigenvalues tends to zero in the vicinity~\footnote{Note
that the positions of the signal and the idler modes slightly change
for increasing densities as a consequence of the interaction-induced
blue-shift of the dispersion.} of $k=k_s$ and $k=k_i=2k_p-k_s$. In
this case, it is easy to see that for wavevectors in the vicinity of
$k_{s,i}$ one has
$\textrm{Re}[E_{k_{s,i}+k}^{(1)}]=\textrm{Re}[E_{k_{s,i}+k}^{(2)}]$
and $\textrm{Re}[E_{k_i-k}^{(1)}]=-\textrm{Re}[E_{k_s+k}^{(1)}]$. In
this region, the eigenvalues can be approximated with the expression
\begin{equation}
{E_k^{(1,2)}}_{\overrightarrow{k\rightarrow k_{s,i}}} \pm
E_{0}+\hbar V_g (k-k_{s,i})-\frac{i\hbar}{2}[\gamma \mp \gamma_0 \pm
\gamma_1 (k-k_{s,i})^2]\,, \label{eq:HDexp}
\end{equation}
where $V_g=\partial_k E_k|_{k_s}/\hbar=\partial_k E_k|_{k_i}/\hbar$
and $\hbar\gamma_0=2g\,|\Phi|^2\lesssim \hbar\gamma$, the equality
holding exactly at threshold $|\Phi|^2=|\Phi_{OPO}|^2$. Following
the procedure already adopted in the previous subsection, we perform
the frequency integration in (\ref{eq:corrkdom}) with the method of
residuals and we expand the function in $k$ around $k_{s,i}$. Then,
by retaining only the dominant contribution (which, as expected,
diverges exactly at threshold $\gamma=\gamma_0$) and using the
relations (\ref{eq:HDexp}), we finally obtain the expressions
\begin{eqnarray}
\bar{G}^{(2)}_{1}(r,t)&\simeq& \frac{2 g}{\hbar}
e^{-(\gamma-\gamma_0)|t|/2}\,\cos(K
r-\frac{E_{0}t}{\hbar})\times\nonumber\\
&\times&\int dk\, \frac{\cos(k R)\,e^{-\gamma_1
k^2|t|/2}}{\gamma-\gamma_0+\gamma_1 k^2}\,, \label{eq:G2kdom1_HD}
\end{eqnarray}
and
\begin{eqnarray}
\bar{G}^{(2)}_{2}(r,t)&\simeq& - \frac{4 \pi g}{\hbar^2 \gamma_0}
e^{-(\gamma-\gamma_0)|t|/2}\times
\nonumber\\
&\times&\mbox{Re}\left\{\int dk\, \frac{e^{i k R-\gamma_1
k^2|t|/2}}{\gamma-\gamma_0+\gamma_1 k^2}\right.\times\nonumber
\\
& \times& \left[(\bar{E}_s+\Delta V_g^{(s)} k +i\hbar\gamma_0/2)\, e^{-i K r+iE_{0}t/\hbar}\right.+\nonumber\\
&+&\left.\left.(\bar{E}_i+\Delta V_g^{(i)}  k +i \hbar\gamma_0/2)\,
e^{i K r-iE_{0}t/\hbar}\right]\right\}\,,\nonumber \\
\label{eq:G2kdom_HD}
\end{eqnarray}
with $R=r-V_g t$, $\bar{E}_{s,i}=\tilde{\epsilon}_{k_{s,i}}\mp E_0$, $K=k_s-k_p$, and
$\Delta V_g^{(s,i)}=V_g-v_g^{(s,i)}$.

This result can be further simplified by introducing the approximations
$\bar{E}_s\simeq \bar{E}_i\simeq 0$ and $\Delta V_g^{(s)}\simeq
\Delta V_g^{(i)} \simeq 0$.
In this case, the anomalous correlation disappears
\begin{eqnarray}
\bar{G}^{(2)}_{2}(r,t)&\simeq& \frac{4\pi g}{\hbar}
e^{-(\gamma-\gamma_0)|t|/2}\cos(K
r-E_{0}t/\hbar)\times\nonumber \\
&\times&\int dk \sin(k R)\frac{e^{-\gamma_1
k^2|t|/2}}{\gamma-\gamma_0+\gamma_1 k^2}=0\,, \label{eq:G2kdom_HD2}
\end{eqnarray}
as the integrand is odd in $k$, and the spatial correlations are
dominated by the normal contribution $\bar{G}^{(2)}\simeq
\bar{G}^{(2)}_1$. The explicit expression of $\bar{G}^{(2)}_1$ is
given in (\ref{eq:G2kdom1_HD}): the Fourier transform of the product
of a Gaussian and a Lorentzian function corresponds, in real space,
to the convolution of an exponential and a gaussian function.

At zero delay $t=0$ [Fig. \ref{fig:HD} (b)], the Gaussian reduces to
a delta-function in space and the correlation signal shows an
exponential decay in space with a characteristic length
\begin{equation}
\ell\simeq \sqrt{\frac{\gamma_1}{\gamma-\gamma_0}}.
\end{equation}
As expected on the basis of general arguments on phase transition,
and previously observed in Monte Carlo
simulations~\cite{carusotto05}, the characteristic length $\ell$
diverges as the OPO threshold is approached $\gamma_0\rightarrow
\gamma$. In contrast to the expression (\ref{eq:G2kdom_LD2}) for the weak intensity case, the zero delay $t=0$ fringe pattern now has a $\cos$ form with a simple bunching peak at $r=0$.

At longer times, the spatial width of the Gaussian grows as
$\sqrt{t}$, so that for short to intermediate distances, the
dependence is dominated by the Gaussian factor. This effect is
clearly visible in Fig. \ref{fig:HD} (c). The overall exponential
decay in time occurs on a characteristic time
\begin{equation}
\tau\simeq\frac{1}{\gamma-\gamma_0}.
\end{equation}
which again diverges as the critical point is approached.

\subsubsection{Non-parametric regime}

The non-parametric regime shown in Fig. \ref{fig:disp} (g,h)
corresponds to the case where the polariton and ghost branches do
not intersect, i.e.
$\tilde{\epsilon}_k+\tilde{\epsilon}_{\bar{k}}\neq 0$ for all $k$.
This regime is generally realized when the frequency of the pump is
very much red-detuned with respect to the renormalized polariton
dispersion and, in our configuration, is fulfilled for pump
intensities above the bistability loop.

In this regime, the Bogolubov eigenmodes can be approximated by the
single particle dispersion blue-shifted by the interaction, i.e.
$E_k\simeq \tilde{\epsilon}_k-i\hbar\gamma/2$ and an equation
formally equivalent to (\ref{eq:G2kdom_LD1}) still holds. However,
since the denominator remains finite for all $k$, no pole can be
identified. However, the region of the polariton dispersion that
minimizes the denominator gives the dominant contribution. In the
case of Fig. \ref{fig:disp} (g), this happens at $k=k_p$. It is
therefore convenient to expand the energy dispersion at second order
in $k-k_p$. Performing this approximation in the denominator, we
obtain the following formula for the correlation function:
\begin{eqnarray}
\bar{G}^{(2)}(r,t)&\simeq&4g(\pi
\frac{g|\Phi|^2}{\tilde{\epsilon}_{k_p}}-2)e^{-\gamma
|t|/2} \times \label{eq:G2kdom1_NOP1}\\
&\times&\mbox{Re}\left\{\int dk \, e^{-ik
r}\frac{e^{i\tilde{\epsilon}_{k+k_p}t/\hbar}}{2\tilde{\epsilon}_{k_p}+2
u_p k^2+i\mbox{Sgn}(t)\hbar\gamma}\right\}\nonumber \,.
\end{eqnarray}
To further simplify this expression, we can expand also the energy
exponents, which leads to
\begin{eqnarray}
\bar{G}^{(2)}(r,t)&\simeq&4 g (\pi
\frac{g|\Phi|^2}{\tilde{\epsilon}_{k_p}}-2)
e^{-\gamma|t|/2}\mbox{Re}\left\{e^{i\tilde{\epsilon}_{k_p}t/\hbar}\times\right.\label{eq:G2kdom1_NOP}\\
&\times&\left.\int d k e^{-i k R} \frac{e^{iu_p k^2 t/\hbar}}{2
\tilde{\epsilon}_{k_p}+2 u_p k^2+i\mbox{Sgn(t)}\hbar\gamma}\right\}
\nonumber \,.
\end{eqnarray}
Here, we have set $R=r-v_g^{(p)} t$, $v_g^{(p)}=\partial_k\tilde{\epsilon}_k|_{k_p}/\hbar$ and
$u_p=\partial^2_k\tilde{\epsilon}|_{k_p}/2$.

The integral over $k$ is of the Fresnel kind and describes the
interference produced at the point $(R,t)$ by the different
$k$-modes with a gapped and quadratic dispersion. For zero
time delay $t=0$, the correlations have a typical anti-bunching character: they are everywhere negative and are strongest at $r=0$. Further away, they monotonically tend to zero
with an exponential law of characteristic length
$\kappa^{-1}=\sqrt{u_p/\tilde{\epsilon}_{k_p}}$ determined by the
gap between the renormalized polariton and ghost
branches~\footnote{It is interesting to note that in the standard
Bogolubov theory of equilibrium systems the ungapped, sonic behavior
of the Bogolubov dispersion leads to a power-law decay of
correlations at $T=0$.}.

The most apparent deviation between the analytical form (\ref{eq:G2kdom1_NOP}) and the numerical result shown Fig.\ref{fig:NOP}(c) consists of a tail in the analytic approximation that extends up to large distances. This has a simple interpretation: the quadratic approximation of the dispersion eliminates all bounds in the group velocity and predicts correlations at any distances. In contrast, the correct dispersion has an upper bound $v_g^{max}$ to the group velocity, which restricts the possible correlations to the $|r|<v_g^{max}\,|t|$ region marked by the thin lines in Fig. \ref{fig:NOP}(a) and (c): this intepretation is confirmed by the much better agreement of the (\ref{eq:G2kdom1_NOP1}) prediction where the group velocity is correctly taken into account.

\section{Two-dimensional and disordered system}
\label{sec:dis}

In this final section we apply our model to the more general case of
a two dimensional inhomogeneous system. In particular, we wish to
investigate how the conclusions of the previous sections are
affected by the presence of exciton and photon disorder.

Realistic system parameters for a GaAs microcavity with $N=2$
quantum wells are used, with $\hbar \Omega_R=3.5$ meV,
$\hbar\gamma_c=\hbar\gamma_x=\hbar\gamma=0.2$ meV, zero
exciton-photon detuning. For the nonlinear interaction constant we
take $g_x=1.5\times 10^{-3}$ meV $\mu$m$^{2}$ and $g_s=0.5\times
10^{-3}$ meV $\mu$m$^{2}$. We consider a configuration where the
pump is orthogonal to the cavity plane, $k_p=0$ and we take
$\hbar\omega_p-\epsilon_c(0)=-3$ meV. This orthogonal pump
configuration is the most suitable one in view of experiments, as it
is least affected by the temporal resolution of the photon
detectors.

\begin{center}
\begin{figure}[htbp]
\includegraphics[width=\columnwidth]{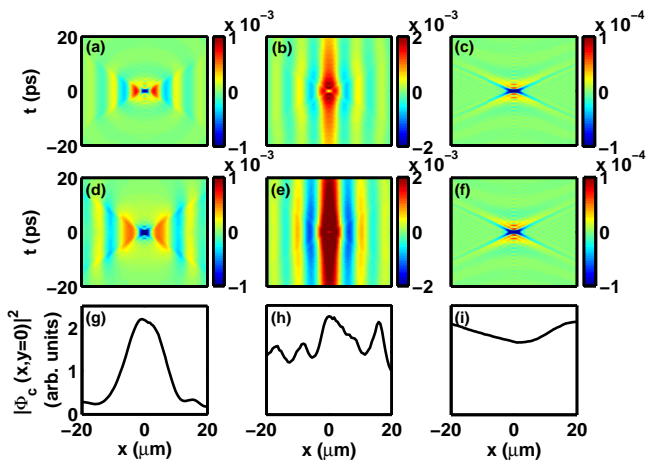}
\caption{Intensity correlation pattern $\bar{G}^{(2)}({\bf r},t)$ as
a function of the spatial coordinate $x$ and of the temporal delay
$t$ along a $y=0$ line for a 2D system. Panels (a-c) are for a
uniform case, while panels (d-f) are for a disordered system. The
pump intensity is varied from a low value (a, d), to a value close
to the OPO threshold (b, e) and to a value well above the
bistability loop (c, f) where parametric processes are forbidden.
Panels (g-i) show the coherent photon field profiles $|\Phi_c(r)|^2$
corresponding to the disordered case of panels (d-f).}
\label{fig:ALLg2}
\end{figure}
\end{center}

The correlation pattern for different values of the pump intensity,
corresponding to the low-density limit, the critical OPO region and
the non-parametric configuration are shown in Fig. \ref{fig:ALLg2}
(a, b, c) for a two dimensional system in the absence of disorder.
All the features discussed in the previous section for the 1D case
are still apparent. In particular, in the low-density case, the
correlations disappear for distances smaller that $|r_{max}(t)|=v_g
t$, $\pm v_g$ being the group velocities of polaritons in the signal
and idler modes, and they decay exponentially. On the other hand, in
the vicinity of the OPO threshold, correlations extend everywhere.
In the non-parametric configuration, correlations are non vanishing
only for distances smaller than $r=v_g^{max} t$, $v_g^{max}$ being
the largest group velocity.

We assume white noise disorder for the exciton field of amplitude
$2$~meV and a gaussian-correlated disorder for the photon field,
with amplitude $U_{dis}^c=0.5$~meV and a correlation length of
$\xi_c=7~\mu$m. The results of the numerical calculations are
summarized in Fig. \ref{fig:ALLg2} (d-i): the profile of the
coherent photon field $|\Phi_c({\bf r})|^2$ in the three cases is
shown in panels (g-i) and the corresponding spatio-temporal patterns
of correlations are shown in panel (d-f). The realization of the
disorder potential is the same for the three values of the pump
intensity.

Even if it is responsible for large density modulations, the
considered disorder is never able to destroy the near-field
correlation pattern. Of course the pattern would eventually
disappear if a much stronger disorder was considered, that is able
to fragment the coherent field in disconnected parts. However, this
latter situation is quite unusual for state-of-the-art GaAs
microcavities.

From the comparison with the corresponding correlation patterns for
a clean system [Fig. \ref{fig:ALLg2} (a-c)], we can still appreciate
a slight modification in the pattern. Disorder is responsible for
these modifications via three main effects: i) the exciton and
photon resonances are broadened; ii) ${\bf k}$-conservation is
broken, which softens the condition for parametric processes; iii)
the group velocity is no longer a well defined concept, so the
contour of the butterfly shape in Fig. \ref{fig:ALLg2} (d) is
smeared out with respect to panel (a).

\section{Conclusions}
\label{sec:concl}

We have developed a formalism to compute the second order spatial
correlation function of polaritons in a planar microcavity. This
quantity directly transfer into the near-field intensity
correlations of the emitted light.

We have computed the spatio-temporal pattern of correlations in
different pumping regimes, ranging from the parametric luminescence
regime, to the critical region just below the parametric oscillation
threshold, and to the strong pumping regime where a large blue-shift
of the polariton modes is able to prevent parametric oscillation.

For each regime, we have identified the key features of the
correlation pattern. In the uniform case, we have compared our
results to the predictions of approximate analytic models which
provide a physical interpretation to the observed patterns. An
orthogonal pump geometry appears as the optimal choice for the
experimental observation as it reduces the required temporal
resolution of photo-detectors. We have verified that the correlation
patterns are not qualitatively modified by a realistic disorder as
long as the coherent polaritons remain delocalized in space.

The conclusions of the present work confirm the expectation that
intensity correlations can be a very powerful tool to study the
dynamics of quantum fields in condensed matter systems. A study of
more complex geometries is presently under way.

\section{Acknowledgements}
We are grateful to Cristiano Ciuti for continuous stimulating
discussions. D. S. acknowledges the financial support from the Swiss
National Foundation (SNF) through the fellowship number
PBELP2-125476.

\section{Appendix}

In this Appendix, we discuss the connection between the photon field inside the cavity and the emitted light. In particular, we show that the calculated in-cavity intensity correlations directly transfer to the near-field correlations of the emitted light.

\begin{center}
\begin{figure}[htbp]
\includegraphics[width=\columnwidth]{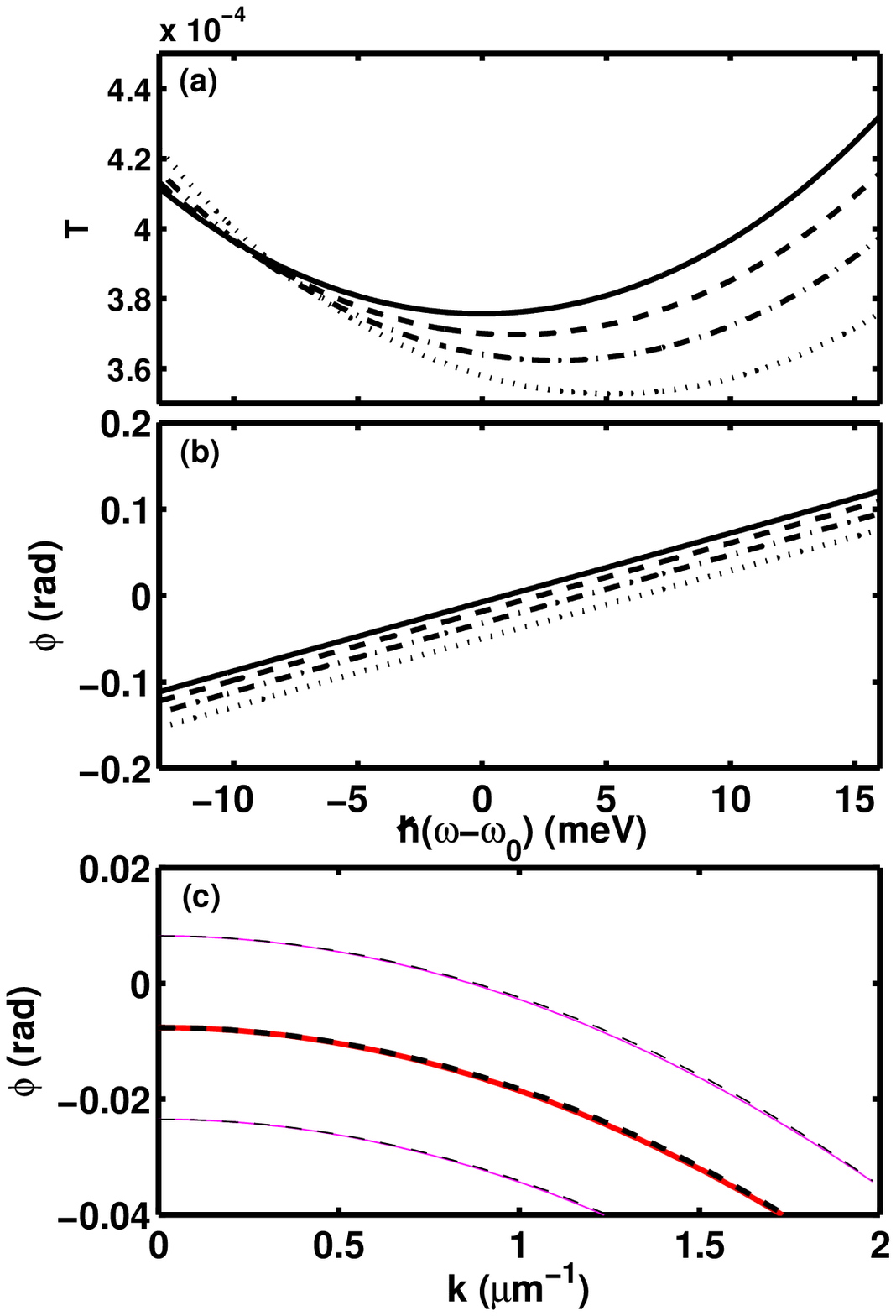}
\caption{Transmittivity (a) and transmission phase (b) across the $\lambda/4$ DBR mirror between cavity and air as a function of the frequency $\omega$, for a given wavevector $k=0$ (solid line), $k=1~\mu$m$^{-1}$ (dashed), $k=1.5~\mu$m$^{-1}$ (dot-dashed) and $k=2~\mu$m$^{-1}$ (dotted).
Parameters are $l_1=65$ nm, $l_2=75$ nm, $n_1=3$, $n_2$=3.5,
$n_c=3.4$, $L_c=290$ nm and $N=22$. The cavity frequency $\omega_0=\pi c/
n_{c}L_c$ is located at the center of the mirror stop band. (c) Transmission phase
across the same DBR mirror as a function of the wavevector $k$, for a give frequency $\omega-\omega_0=0$ (thick solid line) and $\hbar(\omega-\omega_0)=\pm 2$ meV (thin lines). Dashed lines are obtained by fitting the numerical results with Eq. (\ref{eq:phaseT}), which gives $\Delta_z\simeq 140$ nm.} \label{fig:A_tr}
\end{figure}
\end{center}

We consider a planar cavity along the $x-y$ plane with a quantum well placed at $z=0$ and mirrors whose external surface is at $z=\pm z_m$. In Fourier space, the external field $E_{out}({\bf k},\omega)$, resulting from the transmission across the mirror is related to the in-cavity field $E_{in}({\bf k},\omega)$ at the quantum well position via the complex transmission coefficient $\tau({\bf k},\omega)=T^{1/2}e^{i\phi({\bf k},\omega)}$:
\begin{multline}
E_{out}({\bf r},z,t)=\int \frac{d^2{\bf k}}{(2\pi)^2}\int
\frac{d\omega}{2\pi} e^{i({\bf k}\cdot{\bf r}-\omega t)} e^{i k_z(k)
(z-z_m)}\times \nonumber \\
\times \tau({\bf k},\omega) E_{in}({\bf k},\omega)\,.
\label{eq:Eout}
\end{multline}
For simplicity, we have neglected here the ${\bf k}$- and $\omega$-dependence of the transmittivity $T$ as we are resticting to frequencies well within the stop band of the mirror where $T$ is almost constant (see Fig.\ref{fig:A_tr}(a)).

For frequencies close to the $\mathbf{k}=0$ cavity frequency $\omega_0$ and at low wave vectors $\mathbf{k}$, the phase of the transmission coefficient can be accurately approximated by the expansion
\begin{equation}
\phi({\bf
k},\omega)=\phi(0,\omega_0)+\Delta_t(\omega-\omega_0)-\frac{c}{2\omega_0}\Delta_z
k^2\,, \label{eq:phaseT}
\end{equation}
with
\begin{equation}
\Delta_t=\left.\frac{\partial \phi}{\partial
\omega}\right|_{\omega_0}\,
\end{equation}
and
\begin{equation}
\Delta_z=\frac{\omega_0}{c} \left.\frac{\partial^2 \phi}{\partial
k^2}\right|\,.
\end{equation}

The transmitted field thus reduces to
\begin{equation}
E_{out}({\bf r},z,t)\simeq \mathcal{B}\,\int \frac{d^2{\bf k}}{(2\pi)^2} e^{i{\bf
k}\cdot{\bf r}} e^{-i\frac{c}{2\omega_0}(z-\Delta_z)k^2} E_{in}({\bf
k},t-\Delta_t)\,, \label{eq:Eout_bis}
\end{equation}
where we have used the relation (valid in the air)
\begin{equation}
k_z(k)=\sqrt{\omega_0^2/c^2-k^2}\simeq \omega_0/c -c/2\omega_0
k^2\,, \label{eq:kz_vs_k}
\end{equation}
and we have included the transmittivity $T$ and a global phase into the multiplicative constant ${\mathcal B}$.

Without the mirror, the field $E_{out}({\bf r},z,t)$ would be given
by
\begin{equation}
E_{out}({\bf r},z,t)\simeq \int \frac{d^2{\bf k}}{(2\pi)^2} e^{i{\bf
k}\cdot{\bf r}} e^{-i\frac{c}{2\omega_0}z k^2} E_{in}({\bf k},t)\,.
\label{eq:E_nomirror}
\end{equation}
Comparison between Eqs. (\ref{eq:Eout_bis}) and
(\ref{eq:E_nomirror}) shows that the only effect of the mirror on
the external field is to give a time shift of $\Delta_t$ and
a longitudinal space shift $\Delta_z$.

We now demonstrate that the expansion (\ref{eq:phaseT}) holds for a
typical high-quality GaAs microcavity with $N$ DBR's mirrors
\cite{deveaud_specissue}. To this purpose, we employ the standard
transfer matrix approach \cite{savona_ssc} to compute the ${\bf k}$- and $\omega$-dependent complex
transmission coefficient $\tau(\omega,{\bf k})$ across the top mirror placed between the cavity and the
air. We consider the typical case where the DBR's mirrors are
composed by two alternate dielectric layers with index of refraction
$n_1$ and $n_2$ and length $l_1$ and $l_2$, respectively, such as
$l_1/n_1\simeq l_2/n_2\simeq \lambda_0/4$, $\lambda_0=2\pi c/\omega_0$ being the
cavity wavelength. The transfer matrix for the transmission from the
cavity to the air is \cite{savona_ssc}
\begin{equation}
M=T_{a2}T_{21}^{-1}M_{12}^NT_{1c}\,, \label{eq:TMtot}
\end{equation}
where $T_{a2}$, $T_{12}$ and $T_{1c}$ are the matrices describing
the transmission across the interfaces between the air and the
dielectric $2$, between the dielectric $1$ and the dielectric $2$
and between the dielectric $1$ and the cavity, while
\begin{equation}
M_{12}=T_{21}M_{2}T_{12}M_{1}\,,\label{eq:M12}
\end{equation}
is the transfer matrix describing the transmission across the
periodic block formed by the two dielectric layers. Since
\begin{equation}
M\left(\begin{array}{c}1\\\rho\end{array}\right)=\left(\begin{array}{c}\tau\\0\end{array}\right)
\end{equation}
the transmission coefficient is given by
\begin{equation}
\tau=\frac{\mbox{det}(M)}{M_{22}}\,.
\end{equation}

Examples summarized in Fig. \ref{fig:A_tr}. In panel (a), we show the frequency dependent
transmission amplitude $T$, for in-plane wave vectors ranging from
$k=0$ and $k=2~\mu$m$^{-1}$. The relative variation is less than
10\% in the interval of frequencies and wave vectors considered which validates the assumption underlying
(\label{eq:Eout}).
In panel (b) we display the phase of the transmission coefficient as a function of the frequency. For all values of the in-plane wavevector $k$, the linear dependence assumed in Eq. (\ref{eq:phaseT}) is accurately verified with the same $\Delta_t \simeq 5$ fs. This result confirms the validity of the expansion (\ref{eq:phaseT}).
In panel (c), we show the dependence of the phase $\phi$ on the wavevector $k$, for different values of the frequency $\hbar \omega=-2,0,2$ meV. Again the dependence assumed in Eq. (\ref{eq:phaseT}) is verified with the same value of $\Delta_z\sim 140$ nm.

It is important to note that this small spatial shift $\Delta_z$
does not scale with the number $N$ of layers in the DBR mirror, as it could be intuitively supposed. This result originates from the fact that within the stop band the transfer matrix $M_{12}$ (and consequently $M_{12}^N$) has real eigenvalues, and thus does not induce any extensive phase shift.

The results of this appendix show that the only effect of the mirror on the near-field correlation pattern is the following: the light detected outside of the cavity appears to be generated at a slightly earlier time and at a slightly displaced longitudinal position as compared to the quantum well position. In the experiments one has therefore simply to focus the optical detection on this shifted position.


\end{document}